\documentclass[10pt,journal,doublecolumn,singlespacing]{IEEEtran}
\usepackage{url}
\usepackage{setspace}
\usepackage{amsmath}
\usepackage{multicol}
\usepackage{amssymb}
\usepackage[utf8]{inputenc}
\usepackage[T1]{fontenc}
\usepackage[font={small}]{caption}
\usepackage{cite}
\usepackage{enumerate}
\usepackage[square, comma, numbers, sort&compress]{natbib}
\usepackage{subfig}
\usepackage{bm}
\usepackage{multirow}
\usepackage{tabularx}

\ifCLASSINFOpdf
\else
\fi
\ifCLASSINFOpdf
   \usepackage[pdftex]{graphicx}
   \usepackage{epstopdf}
\else
\fi

\begin{document}

\title{\huge Reconfigurable Intelligent Surfaces vs. Relaying:\\ Differences, Similarities, and Performance Comparison}
\author{\normalsize M.~Di~Renzo,~\IEEEmembership{\normalsize Fellow,~IEEE}, 
\normalsize K.~Ntontin,~\IEEEmembership{\normalsize Member,~IEEE},
\normalsize J.~Song,
\normalsize F.~H.~Danufane,
\normalsize X.~Qian,
\normalsize F.~Lazarakis,
\normalsize J.~de~Rosny,
\normalsize D.-T.~Phan-Huy,~\IEEEmembership{\normalsize Member,~IEEE},
\normalsize O.~Simeone,~\IEEEmembership{\normalsize Fellow,~IEEE},
\normalsize R.~Zhang,~\IEEEmembership{\normalsize Fellow,~IEEE},
\normalsize M.~Debbah,~\IEEEmembership{\normalsize Fellow,~IEEE},
\normalsize G.~Lerosey,
\normalsize M.~Fink,
\normalsize S.~Tretyakov,~\IEEEmembership{\normalsize Fellow,~IEEE}, and \normalsize S.~Shamai (Shitz),~\IEEEmembership{\normalsize Fellow,~IEEE} \vspace{-0.75cm}
\thanks{Received Sep. 26, 2019; revised Feb. 21, 2020. M. Di Renzo (corresponding author) is with the Laboratoire des Signaux et Syst\`emes, CNRS, CentraleSup\'elec, Universit\'e Paris-Saclay, 3 rue Joliot Curie, Plateau du Moulon, 91192, Gif-sur-Yvette, France. (e-mail: marco.direnzo@centralesupelec.fr)} }
%
%
%
%
%
%
\maketitle

\section*{Abstract} 
Reconfigurable intelligent surfaces (RISs) have the potential of realizing the emerging concept of smart radio environments by leveraging the unique properties of meta-surfaces. In this article, we discuss the potential applications of RISs in wireless networks that operate at high-frequency bands, e.g., millimeter wave (30-100 GHz) and  sub-millimeter wave (greater than 100 GHz) frequencies. When used in wireless networks, RISs may operate in a manner similar to relays. This paper elaborates on the key differences and similarities between RISs that are configured to operate as anomalous reflectors and relays. In particular, we illustrate numerical results that highlight the spectral efficiency gains of RISs when their size is sufficiently large as compared with the wavelength of the radio waves. In addition, we discuss key open issues that need to be addressed for unlocking the potential benefits of RISs.

\section*{Introduction} 

\subsection{Possible Migration to High-Frequency Bands} 
By 2022, it is expected that the global mobile data traffic will reach a monthly run of 77 exabytes, which corresponds to a 7-fold growth compared with the monthly run of 2017. Such demands may not be accommodated by current cellular standards that utilize only sub-6 GHz frequency bands. A key feature of future wireless networks is hence the potential migration to higher frequencies, e.g., the millimeter (30-100 GHz) and sub-millimeter (above 100 GHz) wave bands \cite{Rappaport_6G}.

Extensive measurements have been conducted at the millimeter wave band and, more recently, the sub-millimeter wave band. These have demonstrated that the use of highly directional steerable antennas enables mobile communication at such high frequencies \cite{Rappaport_6G}. However, millimeter and sub-millimeter wave frequency bands are highly susceptible to blockages from large-size structures, e.g., buildings, on the radio path \cite[Tables 4, 5]{Rappaport_6G}. In addition, millimeter- and sub-millimeter wave signals may be severely attenuated by the presence of small-size objects, e.g., human bodies and foliage.

\vspace{-0.25cm}
\subsection{Relay-Aided Transmission} 
A possible approach for circumventing the unreliability of high-frequency channels is to sense the environment and to identify, on a real-time basis, alternative propagation routes through which the same information-bearing signal can be received. To this end, an established method is the deployment of relays that capitalize on the concept of (distributed) cooperative diversity \cite{MDR_Relays}. The use of relays can effectively turn a single non-line-of-sight (NLOS) link into multiple line-of-sight (LOS) links. This approach requires each relay to be equipped with a dedicated power source and with the necessary front-end circuitry for reception, processing, and re-transmission. For these reasons, the use of relays may result in an increase of the network power consumption and may require a larger capital expenditure for deployment.

In addition, the network spectral efficiency offered by relay-aided systems depends on the duplexing protocol employed for transmission. If a half-duplex (HD) relaying protocol is employed, transmitters and relays are not allowed to transmit concurrently on the same physical resource. This issue can be overcome by employing a full-duplex (FD) relaying protocol, but at the cost of: (i) introducing high \textit{loop-back self-interference} at the relay because of the concurrent transmission and reception of signals; (ii) generating \textit{co-channel interference} at the destination, since relays and transmitters emit different information on the same physical resource; and (iii) increasing the \textit{signal processing complexity} and the \textit{power consumption} of the relays. Relays, therefore, are utilized in an adaptive fashion, depending on channel and interference conditions, for improving the network performance \cite{MDR_Relays}.

\vspace{-0.25cm}
\subsection{Passive Non-Reconfigurable Reflectors} 
When the LOS path is of insufficient quality, another approach to establish alternative routes is through \textit{passive non-reconfigurable specular reflectors}, e.g., dielectric or metallic mirrors \cite{Ismail_Guvenc}. This method for coverage enhancement has the potential benefit of being more cost efficient as compared with relaying, especially in high-frequency bands. However, a main limitation of non-reconfigurable reflectors is that they cannot enable the dynamic shaping of the impinging waves, since their operation cannot be modified after fabrication, i.e., at the time of deployment and operation. Due to the highly dynamic nature of the wireless environment and the nomadic nature of mobile communications, it would be beneficial that such reflectors be capable of adaptively shaping the radio waves based on actual blockage and environmental conditions.

\vspace{-0.25cm}
\subsection{Nearly-Passive Smart Surfaces} 
Propitiously, electromagnetic-based reconfigurable structures that are capable of applying specified transformations to the impinging radio waves do exist and can operate at different frequency bands \cite{Liaskos}, \cite{MDR_Access}. In the literature, these structures are often referred to as large intelligent surfaces, intelligent reflecting surfaces, digitally controllable scatterers, software-controllable surfaces, and \textbf{reconfigurable intelligent surfaces (RISs)}. In this article, we will employ the term RISs in order to highlight their capability of being configurable after deployment. When deployed in wireless networks, RISs have the potential of turning the wireless environment, which is highly probabilistic in nature, into a programmable and partially deterministic space, which is referred to as \textbf{smart (or intelligent) radio environment} \cite{MDR_Eurasip}. 

The aim of this article is to provide an introduction to this topic, with a focus on the differences with relay-aided systems.

\begin{figure}[!t]
	\label{Metasurface_Illustration}
	\centering
	\includegraphics[width=\columnwidth]{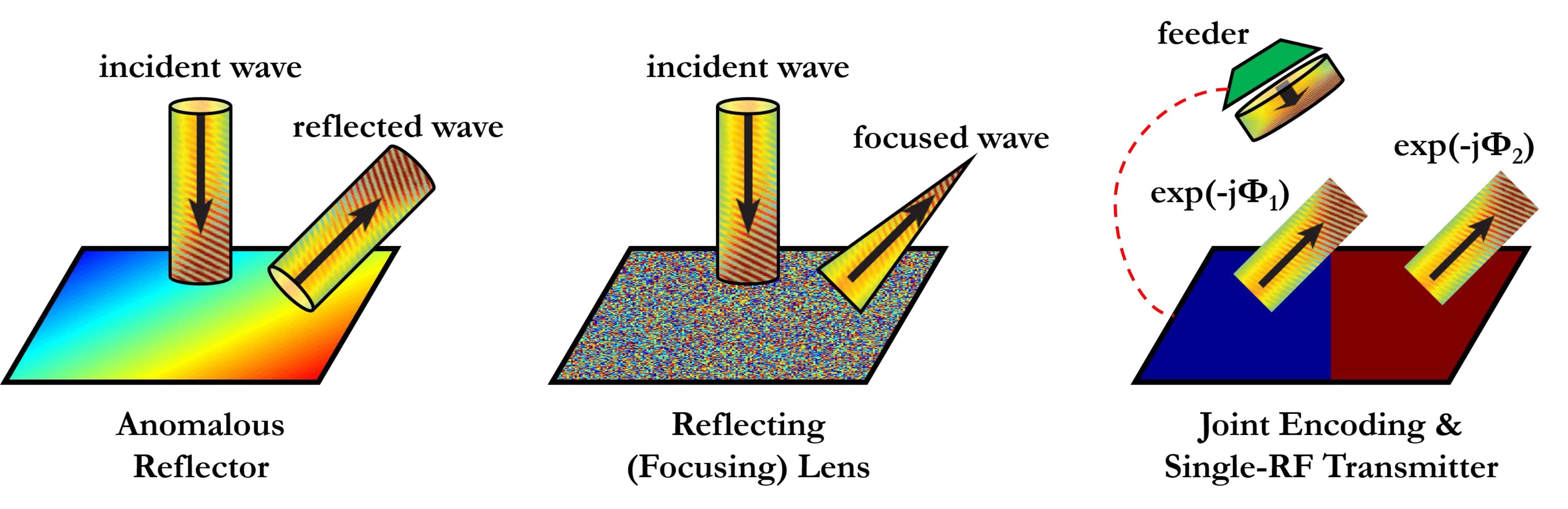}
	\caption{\footnotesize{Possible uses of reconfigurable intelligent surfaces. (i) Anomalous reflection: a radio wave incident at an angle of 90 degrees is reflected towards an angle of 45 degrees. (ii) Focusing lens: a radio wave incident at an angle of 90 degrees is focused (beamforming) towards a specified location in order to maximize the energy at that point. (iii) An RIS illuminated by a feeder reflects two phase-modulated signals by mimicking a two-antenna transmitter, thus encoding information on the reflections of the meta-surface. These functions can be obtained by appropriately configuring the phase response of the RIS (i.e., by optimizing $\Phi(x)$ in \cite[Eq. (4)]{MDR_SPAWC2020}).}}
	\label{Metasurface_Illustration} \vspace{-0.25cm}
\end{figure}    
%
\section*{Reconfigurable Intelligent Surfaces} 
\setcounter{subsection}{0}
\subsection{What is an RIS?} 
An RIS is an artificial surface, made of electromagnetic material, that is capable of customizing the propagation of the radio waves impinging upon it. RISs can be implemented in different ways, including: (i) the realization of large arrays made of inexpensive antennas whose inter-distance is of the order of the wavelength \cite{MIT}; and (ii) the use of meta-material elements whose size and inter-distance is much smaller than the wavelength \cite{Capasso}. In this article, we will focus our attention on the implementation of RISs based on meta-materials, which are referred to as \textbf{meta-surfaces}. 

RISs based on meta-surfaces are very thin -- their thickness is much smaller than the wavelength -- sheets of electromagnetic material that are engineered to possess peculiar properties that cannot be found in naturally occurring materials \cite{Liaskos}-\cite{MDR_Eurasip}. A meta-surface is a sub-wavelength array formed by sub-wavelength metallic or dielectric scattering particles that are referred to as meta-atoms or unit-cells \cite{Liaskos}-\cite{MDR_Eurasip}. It can be described as an electromagnetic discontinuity that is sub-wavelength in thickness, with typical values ranging from $1/10$ to $1/5$ of the wavelength, and is electrically large in transverse size. Its unique properties lie in its capability of shaping the electromagnetic waves in very general ways.

\vspace{-0.25cm}
\subsection{Reconfigurable Meta-Surfaces} 
Meta-surfaces can be either reconfigurable or not. In non-reconfigurable meta-surfaces, the meta-atoms have fixed structural and geometrical arrangements, which result in static interactions with the impinging radio waves that cannot be modified once they are manufactured. In \textbf{reconfigurable meta-surfaces}, the arrangements of the meta-atoms can be modified and programmed based on external stimuli. The reconfigurability can be enabled by electronic phase-changing components, such as semiconductors or graphene, which are used as switches or tunable reactive and resistive elements. They can be either inserted between adjacent meta-atoms or can be used to adjust the properties of individual meta-atoms. As recently demonstrated in \cite{Liaskos}, the wavefront of the radio waves scattered by a meta-surface can be manipulated by controlling the status of the switches, and can be optimized through a central controller based on software-defined networking (SDN) technologies. 

A major difference between static and reconfigurable meta-surfaces lies in their associated power consumption. Static meta-surfaces can be fully passive, since no active electronic circuits are needed. Reconfigurable meta-surfaces can only be \textbf{nearly passive}, since some energy is needed to control the switches, and to receive control signals for configuring them. After the meta-surface is appropriately configured, however, no dedicated power supply is needed for signal transmission. In general, the system to control the meta-atoms and the SDN-based controller are important components of RISs, which affect the rate at which the meta-surfaces are reconfigurable.

\vspace{-0.25cm}
\subsection{Uses of RISs in Wireless Communications} 
In wireless communications and networks, RISs can be employed in multiple ways. In the recent literature, four major uses have been considered as illustrated in Fig. \ref{Metasurface_Illustration}. 

\textbf{Anomalous reflection/transmission} \cite{Capasso}: The RIS is configured in order to reflect or refract the impinging radios waves towards specified \textit{directions} that do not necessarily adhere to the laws of reflection and refraction. The advantage of this application is that the operation of the RIS is independent of the fading channels and the locations of the receivers. The limitation is that, in general, the signal-to-noise-ratio is not maximized and the system capacity is not achieved.

\textbf{Beamforming/focusing} \cite{MIT}: The RIS is configured in order to focus the impinging radio waves towards specified \textit{locations}. The advantage of this application is that the signal-to-noise-ratio is maximized at the locations of interest. The challenge is that, in general, the optimization of the RIS depends on the fading channels and the locations of the receivers. Also, the system capacity is usually not achieved.

\textbf{Joint transmitter/RIS encoding} \cite{Osvaldo}: The RIS is configured in order to optimize the system capacity. The advantage of this application is that the specific status of the meta-atoms is exploited to modulate additional data. The challenge is that, in general, the transmitter and the RIS need to be jointly optimized. In addition, the setup of the RIS depends on the fading channels and the locations of the receivers.

\textbf{Single-RF multi-stream transmitter design} \cite{Wankai_Tx}: This operation is similar to the previous one, with the difference that the transmitter is a simple RF feeder located in close vicinity of the RIS. The feeder emits an unmodulated carrier towards the RIS, which reflects multiple data-modulated signals. This approach is suitable to realize multi-stream transmitters by employing a limited number of (even a single) RF chains. 

Another potential application is the use of RISs for \textbf{increasing the rank of the wireless channel in multiple-antenna systems}. This is discussed in the next section with the aid of a simple example, which is referred to as scattering engineering.

In summary, an RIS can be thought of as a \textbf{multi-function surface} whose use and operation depend on how the meta-atoms are arranged, designed, and optimized. For example, phase gradient meta-surfaces can operate as local phase-gradient reflectors that function as anomalous mirrors, anomalous scatterers, and focusing lenses \cite{MDR_SPAWC2020}. In this article, we are primarily interested in RISs that operate as \textbf{anomalous reflectors}, since they constitute a fundamental element to manipulate the radio waves impinging upon environmental objects and, therefore, to realize smart radio environments.

\section*{Wireless 2.0: Smart Radio Environments} 
\setcounter{subsection}{0}

\subsection{From Adaptation to Control and Programmability} 
From the viewpoint of the communication engineer, the wireless environment is conventionally modeled as an exogenous entity that cannot be controlled, but only adapted to. To this end, communication engineers can only design the transmitters, the receivers, and the transmission protocols in order to achieve the desired performance. Common approaches to capitalize on the properties of the wireless environment and to mitigate its impairments include using multiple antennas, employing complex encoding and decoding algorithms at the end-points of the communication link, and adding additional network infrastructure, e.g., relays, in an attempt to make the transmission of signals more reliable. These solutions, however, may increase the network complexity, the network power consumption, and the network deployment cost \cite{MIT}. 

RISs provide wireless researchers and engineers with a different \textbf{view of the wireless environment}. Since RISs are capable of shaping the wavefront of the radio waves throughout the network, the \textbf{wireless environment can be in principle customized} to suit the system requirements. The wireless environment is not to be treated as a random uncontrollable entity, but rather as part of the network design parameters that are subject to optimization in order to support diverse performance metrics, such as rate, latency, reliability, energy efficiency, privacy, and massive connectivity. The overarching vision consists of coating environmental objects and devices with digitally-controlled RISs, and programming them, through  environmental sensing and SDN-based protocols, for shaping the radio propagation environment and meeting the desired system requirements \cite{Liaskos}, \cite{MDR_Eurasip}.

\begin{figure}[!t]
	\label{Uses of RISs}
	\centering
	\includegraphics[width=1\columnwidth]{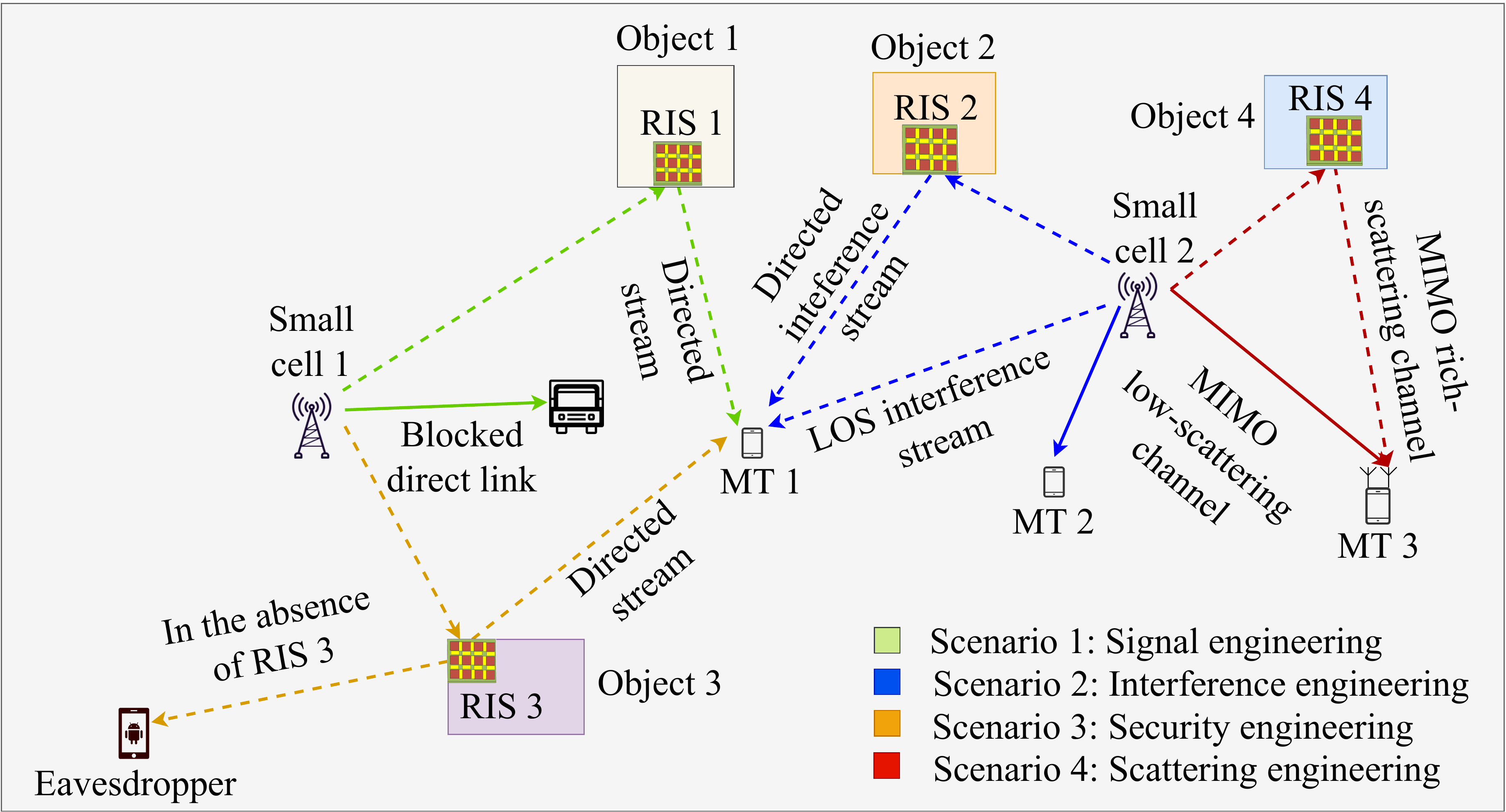}
	\caption{\footnotesize{Example of smart radio environment.}}
	\label{Uses of RISs} \vspace{-0.25cm}
\end{figure}

\vspace{-0.25cm}
\subsection{Illustrative Example of Smart Radio Environment} 
An example of smart radio environment is sketched in Fig.~\ref{Uses of RISs}, where four application scenarios are identified. 

\textbf{Signal engineering}: Assume that small cell 1 wishes to communicate with mobile terminal (MT) 1, but the LOS link is blocked by an object. In this case, small cell 1 redirects the transmitted beam towards RIS 1 that coats object 1, and assists the communication by shaping the incident wave towards MT 1 so that the received signal strength is maximized. 

\textbf{Interference engineering}: While small cell 1 communicates with MT 1, small cell 2 communicates with MT 2. Therefore, an interfering signal reaches MT 1 from small cell 2. To suppress it at MT 1, RIS 2 is programmed to shape the impinging radio wave from small cell 2 towards MT 1 in a way that the two signals are destructively combined at MT 1. 

\textbf{Security engineering}: In the absence of RIS 3, the signal emitted by small cell 1 and intended to MT 1 is reflected from object 3 towards a malicious user that overhears it. To avoid this, RIS 3 is programmed to shape the reflection towards MT 1 so that it is steered away from the malicious user while being decoded more reliably, via diversity combining, at MT 1.

\textbf{Scattering engineering}: The multiple-antenna small cell 2 wishes to convey information to the multiple-antenna MT 3 with the aid of multiple-input multiple-output transmission. The channel between small cell 2 and MT 3 has, however, a low rank (low scattering environment), which negatively affects the attainable data rate. To avoid this issue, small cell 2 directs the signal intended to MT 3 towards RIS 4, which appropriately shapes it so as to create a rich-scattering environment (high rank channel) for high data rate transmission.

\textit{\textbf{From the analysis of these four scenarios, it is apparent that, with the aid of RISs, the propagation of radio waves in wireless networks may be engineered and optimized, at a low complexity, in a way that benefits the network.}}

\section*{Reconfigurable Intelligent Surfaces vs. Relaying} 
\setcounter{subsection}{0}

In this section, we elaborate on differences and similarities between RISs that are employed as anomalous reflectors and relays.  The comparison is made here on a qualitative basis, and is complemented, in the next section, with results that compare RISs and relays on a more quantitative basis.

\vspace{-0.25cm}
\subsection{Hardware Complexity} 
Relays are usually viewed as active devices that need a dedicated power source for operation. They are equipped with active electronic components, such as digital-to-analog converters (DACs) and analog-to-digital converters (ADCs), mixers, power amplifiers for transmission, and low-noise amplifiers for reception. Several electronic components are typically needed for implementing decode-and-forward (DF) and amplify-and-forward (AF) relaying. The deployment of relays may, thus, be costly and power-consuming, especially for realizing multiple-antenna designs at millimeter and sub-millimeter wave frequency bands \cite{Rappaport_6G}. If, in addition, FD relays are used, the complexity is further increased due to the need of eliminating the loop-back self-interference by using tailored antennas and analog/digital signal processing methods. 

In contrast, RISs are composite material layers that are made of metallic or dielectric patches printed on a grounded dielectric substrate. Their configurability is ensured through low-power and low-complexity electronic circuits (switches or varactors) \cite{Capasso}. RISs are envisioned to be of lower complexity than relays, especially at mass production and if realized by using inexpensive large-area electronics, since no dedicated power amplifiers, mixers, and DACs/ADCs are usually required. A prototype of large-size RIS made of 3,720 inexpensive antennas has recently been realized \cite{MIT}. 

\vspace{-0.25cm}
\subsection{Noise} 
The active electronic components used in relays are responsible for the presence of additive noise that negatively affects the performance of conventional relaying protocols. In AF relaying, for example, the noise is amplified at the relays. The impact of additive noise can be mitigated by employing DF relaying, at the expense of decoding and re-encoding (regeneration) the signal at the relays and increasing the signal processing complexity and power consumption. In FD relaying, the impact of residual loop-back self-interference further deteriorates the system performance. 

On the other hand, RISs that behave as anomalous reflectors are not affected by additive noise. However, they may be impaired by phase noises. If they are nearly-passive, in addition, they cannot amplify or regenerate the signals \cite{MDR_Access}.

\vspace{-0.25cm}
\subsection{Spectral Efficiency} 
The spectral efficiency of relay-aided systems depends on the adopted duplexing protocol. Under HD relaying, the achievable rate is generally scaled down by a factor of two, since different physical resources are used for the data emitted by the transmitter and by the relay. The end-to-end signal-to-noise ratio, on the other hand, can be increased by capitalizing on more favorable propagation conditions for the relayed signal, and by optimally combining the direct and relayed signals. Under FD relaying, the achievable rate is not scaled down by a factor of two, but the relay is affected by the residual loop-back self-interference, and the receiver is impaired by the interference generated by the concurrent transmission of the transmitter and the relay. 

RISs that are configured to operate as anomalous reflectors are not subject to the half-duplex constraint and the loop-back self-interference. In addition, the local reflection coefficient of the meta-surface can be designed for optimally combining the signals received from the transmitter and the RIS.

\vspace{-0.25cm}
\subsection{Power Budget} 
Relays require an independent power source for operation, which is used for transmitting the signals (RF power) and for supplying with power their electronic components. 

In contrast, RISs are suitable for nearly passive implementations, since non-reconfigurable meta-surfaces can be realized with fully passive components, and low-power active components (switches or varactors) are needed only for ensuring their reconfigurability. Also, the low-power nature of switches and varactors makes the use of energy harvesting a suitable candidate for realizing close-to-passive implementations. 

In relay-aided systems, it is usually assumed that the total RF power is allocated between the transmitter and the relay, so as to ensure a total power constraint. In RISs, the transmitter uses the total RF power. Also, the power reflected and scattered by the RIS depends on its transmittance, which can be optimized through an appropriate design of the meta-surface \cite{Capasso}. In the ideal case, the total power reflected by an RIS is the same as the total power of the impinging radio wave.

\vspace{-0.25cm}
\subsection{Average Signal-to-Noise Ratio vs. Number of Elements} 
Let us consider a multiple-antenna relay that employs maximum ratio weighting for reception and transmission. If $N$ antennas are used at the relay, the average end-to-end signal-to-noise ratio increases \textit{linearly} with $N$ \cite{MDR_Relays}, \cite{Emil_Relay}. 

On the other hand, the average end-to-end signal-to-noise ratio of an RIS made of $N$ individually tunable antennas (or $N$ reconfigurable meta-surfaces, each of them made of an appropriate number of meta-atoms to realize the desired wave transformations) increases \textit{quadratically} with $N$, while still being subject to the energy conservation principle \cite{MIT}, \cite{Emil_Relay}. Based on existing prototypes for wireless applications, $N$ may be of the order of a few thousands if the RIS is realized by using individually tunable inexpensive antennas \cite{MIT}, and of the order of ten thousands if it is based on meta-surfaces \cite{Wankai_Measurements}.

The different scaling law as a function of $N$ can be understood as follows. In relays, the available power is allocated among the $N$ antennas so that the total power is kept constant. In RISs, in contrast, each constituent antenna or meta-surface reflects, after scaling the received signal by the transmittance and with no noise addition, the same amount of power received from the transmitter. 

It is worth mentioning that, however, the more favorable scaling law as a function on $N$ does not necessarily imply that RISs outperform relays. For a fixed total power constraint, in fact, the path loss as a function of the transmission distance cannot be overlooked. This is discussed next by considering, for ease of exposition and without loss of generality, a free-space propagation model and $N=1$ for both relays and RISs.

\vspace{-0.25cm}
\subsection{Average Signal-to-Noise Ratio vs. Transmission Distance} 
For simplicity and consistency with the numerical results reported in the next section, we consider a two-dimensional space where a  source emits cylindrical radio waves. A relay is assumed to be located at the origin. Likewise, a one-dimensional RIS of length $2L$ is centered at the origin. The distance from the transmitter to the relay/RIS is denoted by $d_{\rm{SR}}$ and the distance from the relay/RIS to the receiver is denoted by $d_{\rm{RD}}$. By using the notation in Table \ref{Table_MathRelays}, the received power as a function of the transmission distance $d$ can be written as ${\left| {E\left( d \right)} \right|^2} \propto {\left( {kd} \right)^{ - 1}}$ \cite{MDR_SPAWC2020}.

Under these assumptions, the end-to-end power received from an AF relay scales with the reciprocal of the product of the transmitter-to-relay distance and the relay-to-receiver distance \cite{MDR_Relays}, i.e., as $\left(k^2 d_{\rm{SR}} d_{\rm{RD}}\right)^{-1}$. When considering the effect of noise, the end-to-end signal-to-noise ratio of both DF and AF relaying scales with the reciprocal of the distance of the weakest of the two paths, i.e., as $\min \left\{ {{(kd_{{\rm{SR}}})}^{ - 1},{(kd_{{\rm{RD}}})}^{ - 1}} \right\}$. 

The total power reflected by an RIS, and, therefore, the scaling law of the received power as a function of the distance, depend on the relation between the geometric size of the RIS, the wavelength of the radio wave, and the relative transmitter-to-RIS and RIS-to-receiver distances. Based on \cite[Sec. III-B]{MDR_SPAWC2020}, two notable regimes are worth of analysis.
\begin{itemize}
\item \textbf{Electrically large RISs}: If the geometric size of the RIS is large enough as compared with the wavelength and the transmission distances ($d_{\rm{SR}}$ and $d_{\rm{RD}}$), the RIS behaves, asymptotically, as an anomalous mirror. In this regime, the power received from the RIS and the end-to-end average signal-to-noise ratio at the receiver scale, as  function of the distance, as ${\left( {{\alpha k d_{{\rm{SR}}}} + {\beta k d_{{\rm{RD}}}}} \right)^{ - 1}}$, where $\alpha$ and $\beta$ depend on the specified angles of incidence and reflection of the radio waves \cite[Eq. (10)]{MDR_SPAWC2020};
\item \textbf{Electrically small RISs}: If the size of the RIS is not large enough as compared with the wavelength and the transmission distances ($d_{\rm{SR}}$ and $d_{\rm{RD}}$), the RIS behaves, asymptotically, as a diffuser. In this regime, the received power and the end-to-end average signal-to-noise ratio at the receiver scale, as a function of the distance, as $4L^2{\left( {d_{{\rm{SR}}} d_{{\rm{RD}}}} \right)^{ - 1}}$ \cite[Eq. (11)]{MDR_SPAWC2020}. This is the same scaling law as for the received power of AF relaying. Notably, the end-to-end average signal-to-noise ratio depends on the length, $2L$, of the RIS.
\end{itemize}

The analysis of electrically large RISs is a relevant case study because of the large geometric size that some implementations of RISs may have. A recent prototype of RIS reported in \cite{Wankai_Measurements}, whose size is $1$ m$^2$ and whose frequency of operation is $10.5$ GHz, is shown to operate in the far-field at distances greater than $70$ m based on analytical formulas and at distances of the order of $28$ m based on experimental measurements. In typical indoor environments, therefore, an RIS of this kind may be viewed as electrically large by transmitters and receivers.

\vspace{-0.25cm}
\subsection{Takeaway Messages from the Comparison} 
Based on the considerations and case studies analyzed in the previous sub-section, it is interesting to compare the scaling laws of RISs and relays as a function of the transmission distance. Let us assume, for simplicity, ${d_{{\rm{SR}}}} = {d_{{\rm{RD}}}} = d_0$, i.e., the RIS/relay is located equidistantly from the transmitter and receiver. Also, let ${M_{\rm{ma}}}$ denote the number of meta-atoms of the RIS  and let $\lambda/D$ with $D>1$ be their inter-distance. Thus, $2L = {M_{\rm{ma}}}\lambda /D$, and the average end-to-end signal-to-noise ratio scales, as a function of the distance, as follows. 
\begin{itemize}
\item Relay-aided transmission: $\propto 1/{d_0}$;
\item Electrically large RIS: $\propto 1/\left( {\alpha {d_0} + \beta {d_0}} \right)$;
\item Electrically small RIS: $\propto 4L^2/d_0^2 \propto M_{\rm{ma}}^2 /d_0^2$.
\end{itemize}

Accordingly, the following conclusions can be drawn:
\begin{itemize}
\item Relay-aided transmission and electrically large RISs (i.e., with a slight abuse of terminology, for short distances $d_0$) offer a similar scaling law as a function of the distance. Since RISs are not subject to the half-duplex constraint and the loop-back self-interference, they have the potential of providing a better rate than relays if, for a fixed size of the RIS, the distances are not too long;
\item Compared with relays, electrically small RISs (i.e., with a slight abuse of terminology, for long distances $d_0$) offer a less favorable scaling law as a function of the distance. However, the average end-to-end signal-to-noise ratio of electrically small RISs scales quadratically with their size, i.e., quadratically with ${M_{\rm{ma}}}$ if $D$ is kept fixed. Thus, a sufficiently large RIS (but still electrically small) has the potential of outperforming relay-aided transmission.
\end{itemize}

Based on these findings, it can be concluded that RIS-aided transmission may outperform relay-aided transmission provided that the size of the RIS is sufficiently large.

\begin{table}[!t] \footnotesize
\centering
\caption{Rate ($R$) for relays and RISs.}
\newcommand{\tabincell}[2]{\begin{tabular}{@{}#1@{}}#2\end{tabular}}
 \begin{tabular}{l|l} \hline
Transmission frequency & $f_c$ \\ \hline
Wavelength & $\lambda$ \\ \hline 
Wave number & $k = {{2\pi } \mathord{\left/ {\vphantom {{2\pi } \lambda }} \right. \kern-\nulldelimiterspace} \lambda }$ \\ \hline 
Electric field (distance $d$)  & $\left| {E\left( d \right)} \right| = {{{E_0}} \mathord{\left/ {\vphantom {{{E_0}} {\sqrt {kd} }}} \right. \kern-\nulldelimiterspace} {\sqrt {kd} }}$  \cite[Eq. (1)]{MDR_SPAWC2020} \\ \hline
Transmit power (RIS) & $P$ \\ \hline 
Transmit power (relay) & $P_R = P/2$ \\ \hline 
Noise power (receiver) & $N_0$ \\ \hline
Self-interference & ${I_{{\rm{S}}}} = 10{N_0}{P_R}$ \\ \hline
HD DF relay & $R = \left( {{1 \mathord{\left/ {\vphantom {1 2}} \right. \kern-\nulldelimiterspace} 2}} \right){\log _2}\left( {1 + \left( {{{{P_R}} \mathord{\left/ {\vphantom {{{P_R}} {{N_0}}}} \right. \kern-\nulldelimiterspace} {{N_0}}}} \right){{\left| {E\left( d \right)} \right|}^2}} \right)$ \\ \hline 
FD DF relay & $R = {\log _2}\left( {1 + \left( {{{{P_R}} \mathord{\left/ {\vphantom {{{P_R}} {\left( {{N_0} + {P_R}{I_{{\rm{LSI}}}}} \right)}}} \right. \kern-\nulldelimiterspace} {\left( {{N_0} + {I_{{\rm{S}}}}} \right)}}} \right){{\left| {E\left( d \right)} \right|}^2}} \right)$ \\ \hline 
Ideal FD DF relay & $R = {\log _2}\left( {1 + \left( {{{{P_R}} \mathord{\left/ {\vphantom {{{P_R}} {{N_0}}}} \right. \kern-\nulldelimiterspace} {{N_0}}}} \right){{\left| {E\left( d \right)} \right|}^2}} \right)$ \\ \hline
RIS - General formula  & $R = {\log _2}\left( {1 + \left( {{{{P}} \mathord{\left/ {\vphantom {{{P}} {{N_0}}}} \right. \kern-\nulldelimiterspace} {{N_0}}}} \right){{\left| {E_{\rm{ris}}\left( d \right)} \right|}^2}} \right)$ \\ \hline
RIS - Mirror (exact) & ${E_{\rm{ris}}\left( d \right)}$ in \cite[Eq. (3)]{MDR_SPAWC2020} \\ \hline
RIS - Mirror (short $d$) & ${E_{\rm{ris}}\left( d \right)}$ in \cite[Eq. (10)]{MDR_SPAWC2020} \\ \hline
RIS - Mirror (long $d$) & ${E_{\rm{ris}}\left( d \right)}$ in  \cite[Eq. (11)]{MDR_SPAWC2020} \\ \hline
RIS - Lens (exact) & ${E_{\rm{ris}}\left( d \right)}$ in \cite[Eq. (3)]{MDR_SPAWC2020} with $\mathcal{P}(x)=0$ \\ \hline
\end{tabular}
\label{Table_MathRelays} \vspace{-0.25cm}
\end{table}

\begin{figure}[!t]
	\centering
	\includegraphics[width=1\columnwidth]{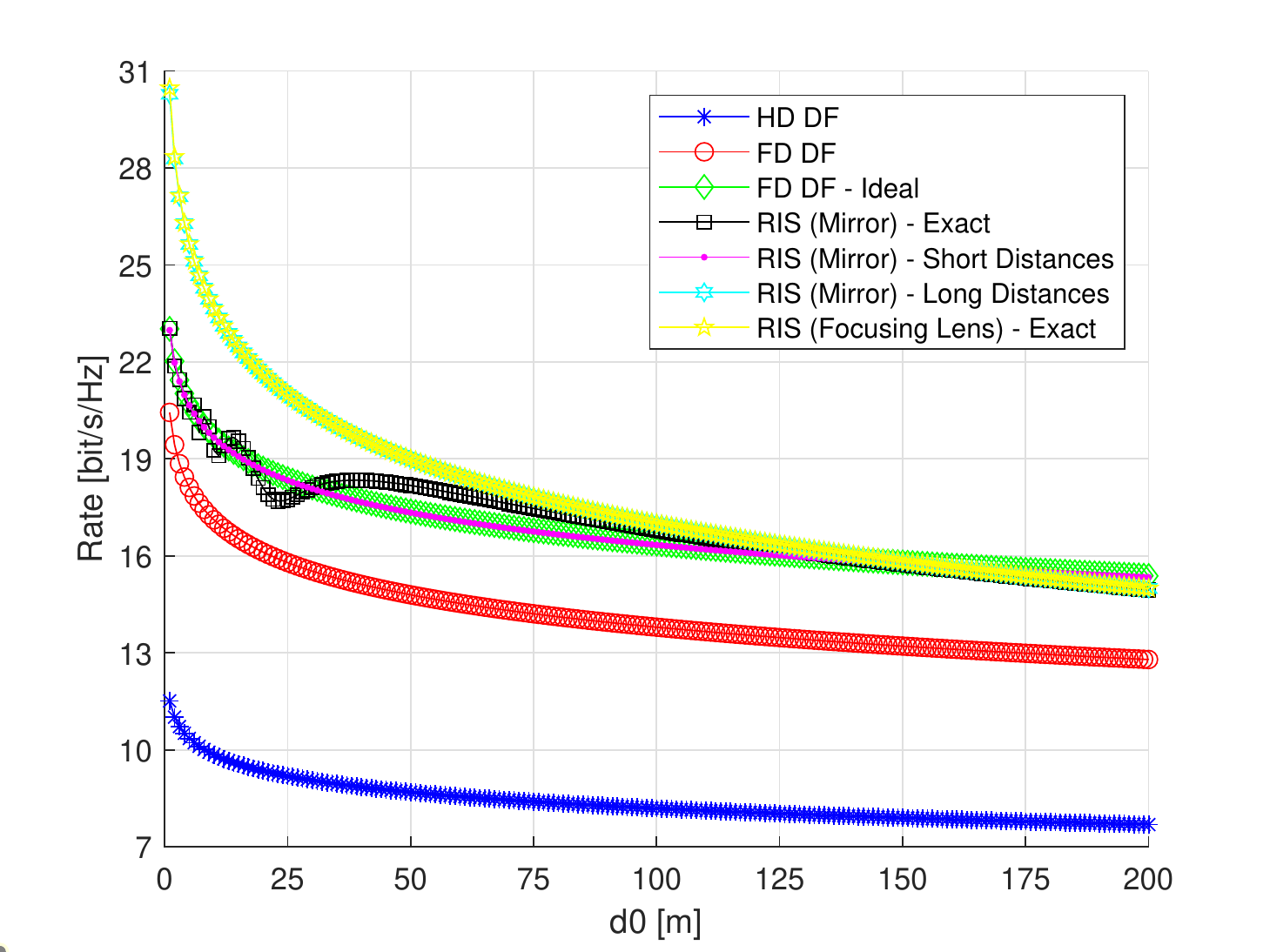}
	\caption{\footnotesize{Data rate of RISs and relays versus the transmission distance.}}
	\label{NumericalResults_Distance} \vspace{-0.25cm}
\end{figure}
\begin{figure}[!t]
	\centering
	\includegraphics[width=\columnwidth]{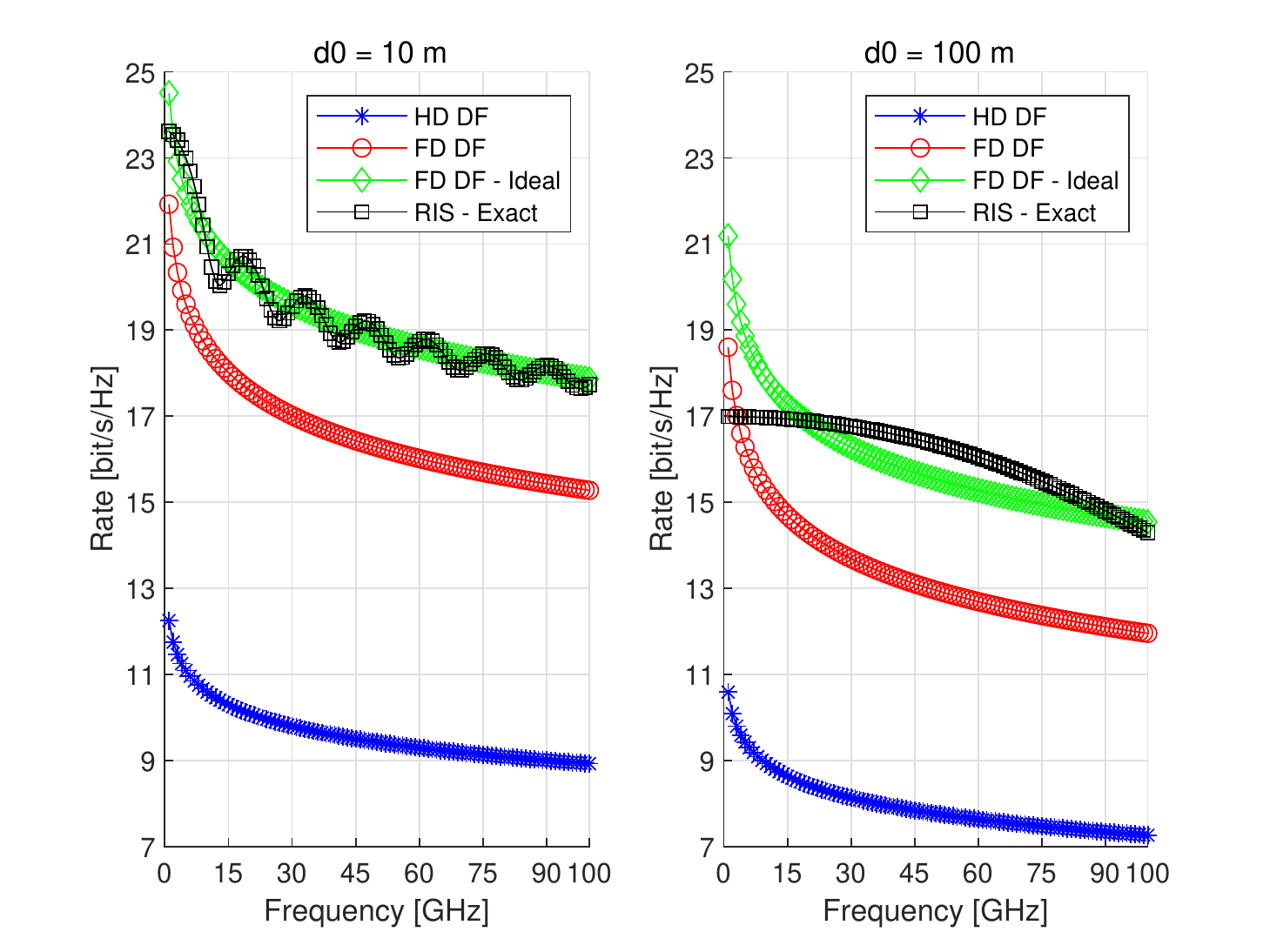}
	\caption{\footnotesize{Data rate of RISs and relays versus the transmission frequency.}}
	\label{NumericalResults_Frequency} \vspace{-0.25cm}
\end{figure}
\begin{figure}[!t]
	\centering
	\includegraphics[width=\columnwidth]{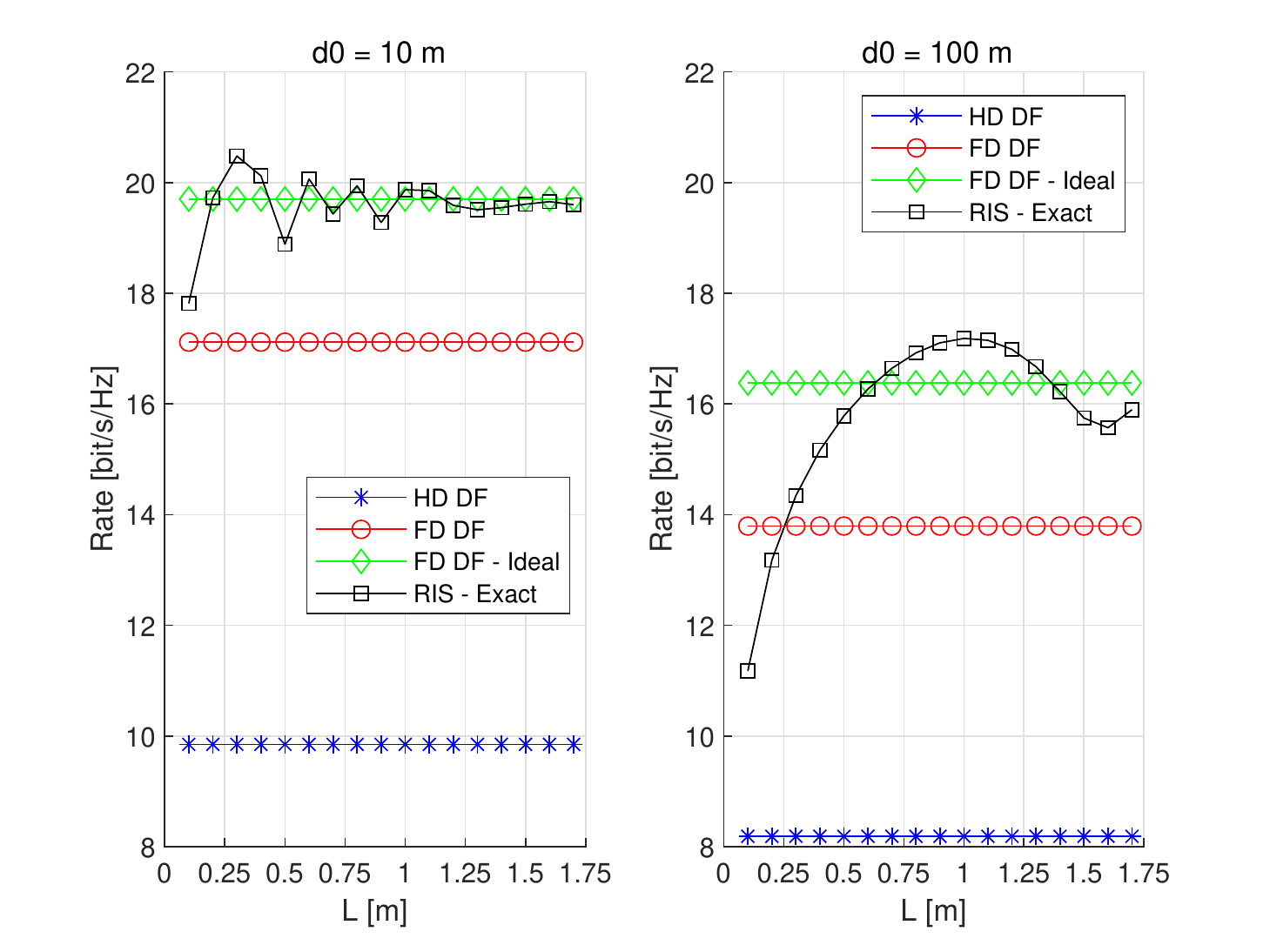}
	\caption{\footnotesize{Data rate of RISs and relays versus the size of the RIS.}}
	\label{NumericalResults_Size} \vspace{-0.25cm}
\end{figure}
%
\section*{Numerical Results} 
\setcounter{subsection}{0}

In this section, we report some numerical illustrations in order to quantitatively compare RISs and relays. For simplicity, we consider a single relay and a single RIS (i.e., $N=1$), and assume that they are located equidistantly from the transmitter and receiver. For the relay, the results are obtained by using the formulas in Table \ref{Table_MathRelays}. For simplicity, only DF relaying is considered, since it provides a better rate than AF relaying, and the performance trends are similar. Table \ref{Table_MathRelays} reports also the rate of an ideal FD relay, in which the residual loop-back self-interference is assumed to be zero. A total power constraint is assumed and, therefore, the total power is equally split between the transmitter and the relay. For the RIS, the intensity of the electric field is obtained from the analytical frameworks in \cite{MDR_SPAWC2020}, as reported in Table \ref{Table_MathRelays}. Without loss of generality, a bi-dimensional system model is assumed as elaborated in \cite[Eq. (1)]{MDR_SPAWC2020}. Therefore, the intensity of the electric field decays with the square root of the distance.

The distance between the transmitter and the relay/RIS, and the relay/RIS and the receiver is denoted by $d_0$. The RIS is modeled as a straight line centered at the origin, which views the transmitter and receiver under an angle of $45$ and $60$ degrees with respect to the normal at the origin, respectively. The total length of the RIS is $2L$. The reflection coefficient of the RIS is chosen as elaborated in \cite[Eqs. (2), (9)]{MDR_SPAWC2020}. Further information about the RIS can be found in \cite{MDR_SPAWC2020}. The signal-to-noise ratio at a distance of $1$ m is ${P \mathord{\left/ {\vphantom {P {{N_0} = X}}} \right. \kern-\nulldelimiterspace} {{N_0} = 114}}$ dB.

\vspace{-0.25cm}
\subsection{RISs vs. Relays as a Function of the Transmission Distance} 
In Fig. \ref{NumericalResults_Distance}, we compare the data rate of an RIS and a relay as a function of the distance $d_0$, by assuming a transmission frequency equal to $f_c = 28$ GHz. For comparison, the RIS is configured to operate as an anomalous reflector and as a focusing lens. This latter case study is discussed next. The RIS is of length $2L=1.5$ m, which corresponds to $140 \lambda$. The exact analytical framework in \cite[Eq. (3)]{MDR_SPAWC2020}, and the approximations for short and long transmission distances in \cite[Eq. (10)]{MDR_SPAWC2020} and \cite[Eq. (11)]{MDR_SPAWC2020}, respectively, are reported. The figure shows that an RIS provides a rate similar to an ideal FD relay without the need of using a power amplifier. This is obtained thanks to the size (effective length) of the RIS. By assuming, for example, that the inter-distance between the meta-atoms of the RIS is in the range $\lambda/5$ and $\lambda/2$, the results in Fig. \ref{NumericalResults_Distance} can be obtained if the number of meta-atoms of the RIS is in the range ${M_{\rm{ma}}} = 700$ and ${M_{\rm{ma}}} = 280$, respectively. The specific implementation depends on the technology employed and the range of directions for which specified anomalous reflection capabilities are needed. It is worth noting that, based on Fig. \ref{NumericalResults_Distance}, the RIS under analysis behaves as an anomalous mirror (i.e., it is viewed as electrically large) for distances $d_0$ up to $25$-$50$ m and as a diffuse scatterer (i.e., it is viewed as electrically small) for distances $d_0$ greater than $75$-$100$ m. Figure \ref{NumericalResults_Distance} shows, in addition, that an ideal FD relay outperforms an RIS for large transmission distances (greater than $150$ m in the considered setup). For long transmission distances, therefore, a larger RIS may be needed for outperforming an ideal FD relay.

\vspace{-0.25cm}
\subsection{RISs: Anomalous Mirrors vs. Focusing Lenses} 
For completeness, Fig. \ref{NumericalResults_Distance} reports the rate of an RIS that is configured to operate as a focusing lens (i.e., a beamformer), as detailed in \cite[Sec. III-C]{MDR_SPAWC2020}. In this latter case, the intensity of the received power scales as a function of the product of the distance between the transmitter and the RIS, and the distance between the RIS and the receiver \cite{Wankai_Measurements}. As expected, Fig. \ref{NumericalResults_Distance} shows that an RIS configured to operate as a focusing lens outperforms, in general, an RIS configured to operate as an anomalous reflector. It is interesting to note that, in the setup of Fig. \ref{NumericalResults_Distance}, an RIS that operates as a focusing lens yields similar rates as the long distance approximation of an RIS that operates as an anomalous reflector (a phase gradient meta-surface \cite[Eq. (9)]{MDR_SPAWC2020}). The price to pay for this performance gain lies in the need of estimating the exact locations of the transmitter and receiver, and in the need of adapting the phases of the RIS to the wireless channels. An anomalous reflector based on a phase gradient meta-surface requires, on the other hand, the knowledge of only the desired directions of incidence and reflection of the radio waves. It is interesting to observe, however, that a sufficiently long RIS that is designed to operate as a simple phase gradient meta-surface is capable of outperforming an ideal FD relay.

\vspace{-0.25cm}
\subsection{RISs vs. Relays as a Function of the Carrier Frequency} 
In Fig. \ref{NumericalResults_Frequency}, we compare the data rate of the RIS and relay as a function of the transmission frequency $f_c$. Two transmission distances are considered, which may be representative of indoor ($d_0=10$ m) and outdoor ($d_0=100$ m) scenarios. The total length of the RIS is $2L=1.5$ m. If $d_0=10$ m, we obtain findings similar to Fig. \ref{NumericalResults_Distance}. If $d_0=100$ m, in contrast, the performance trend is different: If $f_c$ is not large enough (approximately greater than $20$ GHz in the considered example), the length of the RIS is insufficient for outperforming an ideal FD relay. In this case, therefore, an ideal FD relay outperforms an RIS at the price of a higher complexity and power consumption. At higher frequencies, on the other hand, an RIS provides similar rates as an ideal FD relay. This is similar to the findings obtained in Fig. \ref{NumericalResults_Distance}.

\vspace{-0.25cm}
\subsection{RISs vs. Relays as a Function of the Size of the RIS} 
In Fig. \ref{NumericalResults_Size}, we compare the data rate of the RIS and relay as a function of the size of the RIS $L$, by assuming $f_c=28$ GHz. Similar to Fig. \ref{NumericalResults_Frequency}, two transmission distances are analyzed. Once again, we observe that an RIS provides similar rates as an ideal FD relay provided that its is sufficiently (electrically) large as compared with the wavelength $\lambda$. If $d_0 = 100$ m, for example, this holds true if the length of the RIS is of the order of $L=0.5$-$0.75$ m. 

It is worth noting that Figs. \ref{NumericalResults_Distance}-\ref{NumericalResults_Size} show, for short transmission distances, the typical and expected oscillating behavior that is caused by the coherent sum of the many secondary waves, with a different phase, reflected by the RIS \cite[Eq. (3)]{MDR_SPAWC2020}.

\section*{The Road Ahead} 
Theoretical and experimental research on RISs is still at its infancy. Four fundamental and open research issues deserve, in our opinion, more attention than others.

\textbf{Physics-Based Modeling}. Current research on RISs relies on simplified models on how the meta-surfaces shape the impinging radio waves. Hence, there is a compelling need for developing sufficiently accurate but analytically tractable models for the meta-surfaces, whose foundation is to be built on the laws of electromagnetism and physics. For example, RISs are usually modeled as local structures, and, therefore, the spatial coupling among the meta-atoms is ignored. Recent initial results on modeling the mutual coupling of closely-spaced antennas for active surfaces can be found in \cite{Marzetta}.

\textbf{Experimental Validation}. To be accepted by the wireless community, these equivalent models need to be validated through hardware testbeds and empirical measurements. Our analysis reveals that the potential gains and applications of RISs in wireless networks depend on the scaling law of the received power as a function of the distance. There exist, however, only a few experimental results (e.g., \cite{Wankai_Measurements}) that have validated these scaling laws as a function of the size of the RISs, the transmission distances involved, and the specified wave transformations applied by the RISs.

\textbf{Constrained System Design}. The potential gains and applications of RISs in wireless networks depend on their nearly passive implementation. This imposes stringent constraints on the development of efficient signal processing algorithms and communication protocols. The absence of power amplifiers and channel estimation units on the RISs implies, for example, that no channel estimation can be performed at the RISs, and new and efficient (low overhead) protocols need to be developed for acquiring the necessary environmental information for controlling and programming their operation \cite{Cascaded}.

\textbf{Information and Communication Theory}. Conventional information and communication theoretic models applied to wireless networks assume that the system, i.e., the environment, is represented by transition probabilities that are fixed and cannot be optimized. The concept of smart radio environments based on RISs challenges this assumption, allowing the channel states to be included among the degrees of freedom for encoding and modulation. This opens up new venues for system optimization that can provide a better channel capacity, as recently reported in \cite{Osvaldo}.

\section*{Conclusions} 
RISs are an emerging and little understood technology with several applications in wireless networks. In this article, we have discussed the differences and similarities between relays and RISs that are configured to operate as anomalous reflectors. This article complements the numerical study performed in \cite{Emil_Relay}, where the authors compare the power consumption and the energy efficiency of DF relays against RISs that operate as focusing lenses. With the aid of simple scaling laws and numerical simulations, we have provided arguments showing that sufficiently large RISs can outperform relay-aided systems in terms of data rate, while reducing the implementation complexity. The obtained results unveil the advantages and limitations, as compared with relays, of employing RISs that operate as anomalous reflectors in wireless networks.

\bibliographystyle{IEEEtran}

\end{document}